\documentclass[12pt,dvipdfmx]{article}

\catcode`\@=11
\@addtoreset{equation}{section}

\evensidemargin \oddsidemargin

\usepackage[dvipdfmx]{graphicx,hyperref}
\hypersetup{
setpagesize=false,
bookmarksnumbered=true,
bookmarksopen=true,
colorlinks=true,
linkcolor=black,
citecolor=black,
urlcolor=black,}

\usepackage{mathrsfs,amsbsy,amssymb,latexsym,amsfonts,amsmath,amssymb,}

 \usepackage{bm}

\usepackage[dvipdfmx]{graphicx}

\usepackage{color}

\usepackage{comment}

\usepackage{cite}

\newcommand\CN{\mathcal{N}}
\newcommand\CW{\mathcal{W}}
\newcommand\Fs{F\hspace{-2.7mm}/\:}

\newcommand\Ab{\boldsymbol{A}}

\newcommand\Fb{\boldsymbol{F}}

\newcommand\pa{\partial}

\newcommand\Phib{\boldsymbol{\Phi}}
\newcommand\CWb{{\boldsymbol{\mathcal{W}}}}
\newcommand\Fsb{{\boldsymbol{{F\hspace{-2.9mm}/\,}}}}

\newcommand\nn{\nonumber}

\newcommand\adss[2]{AdS$_{#1}\times$S$^{#2}$}

\newcommand\tr{{\rm tr}\,}

\renewcommand\d{{\rm d}}

\newcommand\bref[1]{(\ref{#1})}

\begin{document}

\vspace*{3cm}

\begin{center}
{\Large \bf 
Supersymmetric Non-abelian DBI Equations
 \\[0.50cm]
from
Open Pure Spinor Superstring
}
\vspace*{3cm}\\
{\large
Ryota Fujii,
Sota Hanazawa$^1$,
Hiraki Kanehisa,
and
Makoto Sakaguchi\footnote{makoto.sakaguchi.phys@vc.ibaraki.ac.jp}
}
\end{center}
\vspace*{1.0cm}
\begin{center}

Department of Physics, Ibaraki University, Mito 310-8512, Japan
\\[0.3cm]
$^1$Tsuchiura Nihon University High School,
Tsuchiura 
300-0826,
Japan
\end{center}

\vspace{2cm}

\begin{abstract}
The BRST invariance of the open pure spinor superstring is examined
in the presence of background superfields
on a D$p$-brane.
We note that the background superfields
introduced in this paper depend on boundary fermions.
The BRST invariance leads
 to supersymmetric Dirac-Born-Infeld (DBI) equations
for background superfields depending on boundary fermions
as well as 
boundary conditions on spacetime coordinates.
After quantizing boundary fermions,
background superfields are promoted to non-abelian ones.
As a result, we obtain
the supersymmetric non-abelian DBI equations
from
the supersymmetric DBI equations depending on boundary fermions.
It is shown that these non-abelian DBI equations reduce
to the super-Yang-Mills equations in the limit $\alpha'\to0$.
We also show the nilpotency of the BRST transformation of boundary fermions.

\end{abstract}

\thispagestyle{empty}
\setcounter{page}{0}

\newpage

\tableofcontents

\setcounter{footnote}{0}

\section{Introduction}

Non-abelian Dirac-Born-Infeld (DBI) theory is known as a theory which  describes,
along with the Wess-Zumino action,
 the low-energy dynamics
of an open string attaching to coincident D$p$-branes.
 In the limit $\alpha'\to 0$, it reduces to the Yang-Mills theory. 
 The supersymmetric generalization was achieved in the Green-Schwarz formalism.
By examining the classical $\kappa$-invariance of an open superstring
in an abelian background,
the supersymmetric abelian DBI equations of motion are derived in \cite{GS open background}.
Furthermore, in \cite{boundary fermion} the non-abelian generalization was discussed 
introducing 
boundary fermions \cite{SFT boundary fermion} which correspond to Chan-Paton factors describing coincident D-branes after quantizing boundary fermions.

In \cite{PS open background}, the BRST charge conservation of an open pure spinor superstring in the 
presence of background superfields 
is shown to imply the supersymmetric abelian  DBI 
equations on a D9-brane. This approach also analyzed non-abelian backgrounds up to the second order in boundary fermions.
Recently, in \cite{HS DBI}, 
supersymmetric abelian DBI equations on a D$p$-brane
was obtained from the BRST invariance of the open pure spinor superstring.
The main subject of this paper is to introduce boundary fermions in this model
and to derive the 
supersymmetric 
non-abelian 
DBI equations on coincident D$p$-branes.

We add two terms to the open pure spinor superstring action $S_0$.
One is $S_b$ which is needed to preserve half of the 32 supersymmetry of $S_0$
without imposing any boundary condition,
and the other is $V$ which is the coupling to the background superfields on a D$p$-brane.
In this paper, the background superfields are functions of boundary fermions.
As in \cite{HS DBI},
we demand the BRST invariance of $S_0+S_b+V$,
and solving the condition for it we obtain the supersymmetric DBI equations with boundary fermions
as well as boundary conditions on the spacetime coordinates.
By quantizing these boundary fermions, the supersymmetric DBI equations  are promoted to 
the non-abelian ones.

\medskip

This paper is organized as follows.
After introducing the open pure spinor superstring action $S_0$,
we present a boundary action $S_b$
which preserves half of 32 supersymmetries of $S_0$,
without 
imposing any boundary condition
for a D$p$-brane.
In \S 3,
we introduce a background superfield coupling $V$
which is composed of background superfields depending on boundary fermions.
We solve conditions for the BRST invariance of $S_0+S_b+V$,
and obtain 
supersymmetric DBI equations for background superfields
depending on boundary fermions
in \S4.
In \S5,
by quantizing boundary fermions, 
background superfields depending on boundary fermions are promoted to 
non-abelian ones.
As a result,
we obtain 
 supersymmetric non-abelian DBI equations.
In the limit $\alpha'\to 0$, these equations are shown to reduce to super-Yang-Mills (SYM) equations.
The last section is devoted to a summary and discussions.
In Appendix \ref{sec:nilpotency}, we prove the nilpotency of the BRST transformation of boundary fermions,
which are determined by the BRST invariance of $S_0+S_b+V$.

\section{Open Pure Spinor Superstring
and Boundary Action}  

The action of the pure spinor superstring \cite{PS}
in conformal gauge  is given as
\begin{align}
S_0=&\frac{1}{\pi\alpha'}\int \d z\d \bar z \Big[
\frac{1}{2}\pa x^m\bar\pa x_m
+p\bar\pa\theta
+\hat p\pa\hat\theta
+w\bar\pa\lambda
+\hat w\pa\hat\lambda
\Big]
\end{align}
where $x^m\, (m=0,1,\dots,9)$ are spacetime coordinates,
and $(\theta^\alpha,\hat \theta^\alpha)$ ($\alpha=1,\dots,16$)
are a pair of ten-dimensional Majorana-Weyl spinors.
A pair of bosonic ghosts
$(\lambda^{\alpha},\hat \lambda^{\alpha})$
satisfy pure spinor constraints
$\lambda\gamma^m\lambda=\hat\lambda\gamma^m\hat\lambda=0$.
The $(p_\alpha,\hat p_\alpha)$ and $(w_\alpha,\hat w_\alpha)$
are momenta conjugate to $(\theta^\alpha,\hat \theta^\alpha)$ and $(\lambda^{\alpha},\hat \lambda^{\alpha})$
respectively.
Our convention for the hermite conjugate of $\psi_1\psi_2$ is $(\psi_1\psi_2)^\dag=-\psi_2^\dag \psi_1^\dag$
when $\psi_i$ is Grassmann-odd.
We define worldsheet derivatives as $\pa=\pa_\tau+\pa_\sigma$ and $\bar\pa=\pa_\tau-\pa_\sigma$.
Variation of $S_0$ leads to equations of motion
\begin{align}
\pa\bar\pa x^m=0,~~
\bar\pa\theta
=\bar\pa p
=\bar\pa w
=\bar\pa\lambda
=0,~~
\pa\hat \theta
=\pa \hat p
=\pa \hat w
=\pa\hat \lambda
=0.
\label{eqn:eom}
\end{align}

The action $S_0$ is invariant under the
ten-dimensional $\CN=2$ supersymmetry transformation
with a pair of sixteen parameters, $\epsilon^\alpha$ and $\hat \epsilon^\alpha$,
\begin{align}
\delta\theta^\alpha=\,&\epsilon^\alpha,~~
\delta\hat\theta^\alpha=\hat\epsilon^\alpha,~~
\delta x^m=\frac{1}{2}\theta\gamma^m \epsilon
+\frac{1}{2}\hat \theta\gamma^m \hat\epsilon
,~~
\delta\lambda^\alpha=\delta\hat\lambda^\alpha=0,~~
\delta w_\alpha=\delta \hat w_\alpha=0,~~
\nn\\
\delta p_\alpha=\,&\frac{1}{2}\pa x^m (\epsilon\gamma_m)_\alpha
-\frac{1}{8}\epsilon\gamma^m\theta\, (\pa\theta\gamma_m)_\alpha,~~
\delta \hat p_\alpha=\frac{1}{2}\bar\pa x^m (\hat \epsilon\gamma_m)_\alpha
-\frac{1}{8}\hat \epsilon\gamma^m\hat \theta\, (\bar\pa\hat \theta\gamma_m)_\alpha.
\end{align}
A key relation to see the invariance is the Fierz identity
\begin{align}
(\gamma_m)_{\alpha(\beta}(\gamma^m)_{\gamma\delta)}=0,~~~
(\gamma^m)^{\alpha(\beta}(\gamma_m)^{\gamma\delta)}=0,
\label{eqn:Fierz}
\end{align}
which is frequently  used in this paper.
For later use, we introduce $\epsilon$ supersymmetry invariants
\begin{align}
\Pi^{m}=\pa x^m+\frac{1}{2}\theta\gamma^m\pa\theta,~~
d_\alpha=p_\alpha
-\frac{1}{2}\pa x^m(\theta\gamma_m)_\alpha
-\frac{1}{8}\theta\gamma^m \pa\theta\,(\theta\gamma_m)_\alpha,
\label{eqn:d}
\end{align}
and $\hat \epsilon$ supersymmetry invariants
\begin{align}
\hat \Pi^{m}=\bar\pa x^m+\frac{1}{2}\hat \theta\gamma^m\bar\pa\hat \theta,~~
\hat d_\alpha=\hat p_\alpha
-\frac{1}{2}\bar\pa x^m(\hat \theta\gamma_m)_\alpha
-\frac{1}{8}\hat \theta\gamma^m \bar\pa\hat \theta\,(\hat \theta\gamma_m)_\alpha.
\label{eqn:hat d}
\end{align}
For an open string,
 we are left with a surface term,
\begin{align}
\delta S_0=&\frac{-1}{2\pi\alpha'}\int\d\tau\Big[
\frac{1}{2}(\theta\gamma_m\epsilon-\hat \theta\gamma_m\hat \epsilon)\dot x^m
-\frac{1}{12}\epsilon\gamma^m\theta\,\theta\gamma_m\dot\theta
+\frac{1}{12}\hat \epsilon\gamma^m\hat \theta\,\hat \theta\gamma_m\dot{\hat \theta}
\Big]\Big|_{\sigma=\sigma_*},
\label{susy variation S0}
\end{align}
where $\dot A$
denotes $\pa_\tau A$. 
To eliminate this surface term, we may impose a boundary condition
at the boundary $\sigma=\sigma_*$.
Understanding that the boundary is always at $\sigma =\sigma^*$,
we omit ``$|_{\sigma=\sigma^*}$'' in the following.
In the presence of a D$p$-brane,
we usually impose 
Neumann boundary condition $\pa_\sigma x^\mu =0$ for $\mu=0,1,\cdots,p$
and
Dirichlet boundary condition $\delta x^i =0$ for $i=p+1,\cdots,9$.
For fermionic variables $\theta$ and $\epsilon$, we impose
\begin{align}
\hat\theta=&\gamma^{1\cdots p}\theta,~~~
\hat\epsilon=\gamma^{1\cdots p}\epsilon.
\label{epsilon relation}
\end{align}
For consistency, $p$ must be odd for the type IIB string, while $p$ must be even for the type  IIA string.
It is straightforward to see that these eliminate \bref{susy variation S0}.

\medskip

\subsection{Supersymmetry and Boundary Action}

In this paper, 
we will add a boundary action $S_b$
instead of imposing boundary conditions on fields,
and require that $S_0+S_b$ 
should be invariant under a half of supersymmetries
satisfying the latter equation in \bref{epsilon relation}.
Such supersymmetry transformations 
generated by $\epsilon_+$ are
\begin{align}
\delta_+\theta_+^\alpha=&\epsilon_+,~~
\delta_+\theta_-^\alpha=0,~~
\delta_+ x^\mu=\frac{1}{2}\theta_+\gamma^\mu\epsilon_+,~~
\delta_+ x^i=\frac{1}{2}\theta_-\gamma^i\epsilon_+,~~
\delta_+ \lambda_\pm^\alpha=\delta_+ w^\pm_\alpha=0,
\label{epsilon_+ susy}
\end{align}
where we 
have defined the following objects
\begin{align}
\theta_\pm^\alpha=\frac{1}{\sqrt{2}}(\hat\theta\pm \gamma^{1\cdots p}\theta)^\alpha,~~~~
\epsilon_\pm^\alpha=\frac{1}{\sqrt{2}}(\hat\epsilon \pm \gamma^{1\cdots p}\epsilon)^\alpha,~~~
\lambda_\pm^\alpha \equiv& \frac{1}{\sqrt{2}}(\hat \lambda\pm \gamma^{1\cdots p}\lambda)^\alpha,
\end{align}
and similarly
\begin{align}
p_\alpha^\pm\equiv & \sqrt{2}  (\hat p\pm p \gamma^{p\cdots 1})_\alpha,~~&
d^\pm_\alpha\equiv & \sqrt{2}(\hat d \pm d\gamma^{p\cdots 1})_\alpha,
&w^\pm_\alpha\equiv &\sqrt{2}(\hat w\pm w \gamma^{p\cdots 1})_\alpha.
\end{align}
The $\epsilon_+$ supersymmetry transformation of $S_0$ is found to be
\begin{align}
\delta_+ S_0=\frac{1}{2\pi\alpha'}\int\d\tau\bigg[&
\frac{1}{2}\theta_-\gamma_\mu\epsilon_+ \,\dot x^\mu
+\frac{1}{2}\theta_+\gamma_i\epsilon_+ \,\dot x^i
-\frac{1}{8}\epsilon_+ \gamma^m \theta_+ \, \theta_-\gamma_m \dot \theta_+
-\frac{1}{24}\epsilon_+\gamma^m\theta_-\, \theta_-\gamma_m \dot\theta_-
\bigg].
\label{eqn:delta+ S_0}
\end{align}

The boundary action $S_b$
which satisfies $\delta_+(S_0+S_b)=0$
is given in \cite{HS DBI}
as
\begin{align}
S_b=\frac{1}{2\pi\alpha'}\int \d \tau\bigg[&
\frac{1}{2}\Pi^\mu_+\theta_+\gamma_\mu \theta_-
-\frac{1}{2}x^i \theta_+\gamma_i\dot\theta_+
-\frac{1}{8}{\theta}_+\gamma^m\theta_-\,\theta_+\gamma_m\dot\theta_+
\nn\\&
+\frac{1}{24}{\theta}_+\gamma^m\theta_-\,\theta_-\gamma_m\dot\theta_-
+\frac{1}{2}\Delta^+\theta_-
+\frac{1}{2}w^+\lambda_-
+y_i \Pi^i_-
\bigg].
\label{eqn:S_b}
\end{align}
We have introduced the following objects
\begin{align}
\Pi^\mu_+\equiv\,&
\frac{1}{2}(\hat\Pi^{\mu}+\Pi^{\mu})-\frac{1}{2}\theta_-\gamma^\mu\dot\theta_-
=\dot x^\mu+\frac{1}{2}\theta_+\gamma^\mu \dot\theta_+,
\\
\Pi^i_+\equiv\,&
\frac{	1}{2}(\hat\Pi^{i}+\Pi^{i})-\frac{1}{2}\theta_-\gamma^i\dot\theta_+
=\dot x^i +\frac{1}{2}\theta_+\gamma^i \dot\theta_-,
\\
\Pi^\mu_-\equiv\,&
\frac{1}{2}(\hat \Pi^\mu-\Pi^\mu)-\frac{1}{2}\theta_-\gamma^\mu \dot\theta_+
=-x'^\mu +\frac{1}{2}\theta_+\gamma^\mu\dot\theta_-,
\\
\Pi^i_-\equiv\,&
\frac{1}{2}(\hat \Pi^i-\Pi^i)-\frac{1}{2}\theta_-\gamma^\mu \dot\theta_-
=-x'^i +\frac{1}{2}\theta_+\gamma^i\dot\theta_+,
\label{Pi+}
\\
\Delta^+_\alpha\equiv&~d^+_\alpha
+\Pi^\mu_-(\theta_-\gamma_\mu)_\alpha
+\Pi^i_+(\theta_-\gamma_i)_\alpha
+\frac{1}{2}\theta_-\gamma^m\dot\theta_+ (\theta_-\gamma_m)_\alpha,
\\
y^i\equiv \,&x^i+\frac{1}{2} \theta_+\gamma^i \theta_-.
\end{align}
These
are invariant under the
$\epsilon_+$ supersymmetry transformations
\bref{epsilon_+ susy}
subject to
\begin{align}
\pa_\sigma \theta_\pm=-\pa_\tau \theta_\mp,~~~
\pa_\sigma \lambda_\pm=-\pa_\tau \lambda_\mp,
\label{eqn:eom theta}
\end{align}
which are consistent with the bulk equations of motion \bref{eqn:eom}.
In the following, we will use \bref{eqn:eom theta} frequently on the boundary.
Note that 
 the last three terms in \bref{eqn:S_b}\footnote{We set constants $c_1$ and $c_2$ in \cite{PS open background, HS DBI} as $c_1=c_2=1$ by redefinitions of background superfields.} are invariant separately under the $\epsilon_+$ supersymmetry.
These three terms are required by the BRST invariance in the absence of background superfields, as will be seen below.

\subsection{BRST-variation of $S_0+S_b$}

$S_0$ is invariant under a pair of BRST variations $\delta_B^1$ and $\delta_B^2$
separately.
In the presence of the boundary, the BRST current conservation
may leave $S_0$ invariant under
the BRST variation $\delta_B$  which is the sum of them 
$\delta_B^1+\delta_B^2$.
The BRST transformation law
 is given as follows
\begin{align}
\delta_ B x^m
=&\frac{1}{2}\hat\lambda\gamma^m\hat \theta
+\frac{1}{2}\lambda\gamma^m\theta,~~~
\delta_B\theta=\lambda,~~~
\delta_B\hat\theta=\hat\lambda,~~~
\delta_B \lambda=\delta_B \hat\lambda=0,~~~
\delta_B\omega=d,~~~
\delta_B \hat\omega=\hat d,~~~
\nn\\
\delta_B p_\alpha=&
-\frac{1}{2}\pa x^m(\lambda\gamma_m)_\alpha
+\frac{3}{8} \theta\gamma^m\lambda
(\pa\theta\gamma_m)_\alpha
+\frac{1}{8}\theta\gamma^m\pa\lambda(\theta\gamma_m)_\alpha,
\\
\delta_B \hat p_\alpha=&
-\frac{1}{2}\bar\pa x^m(\hat \lambda\gamma_m)_\alpha
+\frac{3}{8} \hat \theta\gamma^m\hat \lambda
(\bar \pa\hat \theta\gamma_m)_\alpha
+\frac{1}{8}\hat \theta\gamma^m\bar\pa\hat \lambda(\hat \theta\gamma_m)_\alpha,
\nn
\end{align}
where $d$ and $\hat d$ are given in \bref{eqn:d} and \bref{eqn:hat d}, respectively.
We note $\delta_B$ acts from the left.
The $S_0$ is found to be BRST invariant
up to a surface term
\begin{align}
\delta_B S_0=
\frac{1}{2\pi\alpha'}
\int\d\tau
\bigg[&
\frac{1}{2}\Pi^\mu_+(\lambda_+\gamma_\mu\theta_-
+\lambda_-\gamma_\mu\theta_+)
+\frac{1}{2}\Pi^i_+(\lambda_+\gamma_i\theta_++\lambda_-\gamma_i\theta_-)
\nn\\&
-\frac{1}{4}\theta_+\gamma^\mu \dot\theta_+
(\lambda_+\gamma_\mu\theta_-+\lambda_-\gamma_\mu\theta_+)
-\frac{1}{4}\theta_+\gamma^i \dot\theta_-
(\lambda_+\gamma_i\theta_++\lambda_-\gamma_i\theta_-)
\nn\\&
+\frac{1}{8}(\theta_+\gamma^m\dot\theta_++\theta_-\gamma^m\dot\theta_-)
(\lambda_+\gamma_m\theta_-+\lambda_-\gamma_m\theta_+)
\nn\\&
+\frac{1}{8}(\theta_+\gamma^m\dot\theta_-+\theta_-\gamma^m\dot\theta_+)
(\lambda_+\gamma_m\theta_++\lambda_-\gamma_m\theta_-)
\bigg].
\label{eqn:delta_Q S_0}
\end{align}
The BRST variation $\delta_B S_b$ is found to be
\begin{align}
\delta_B S_b=&\frac{1}{2\pi\alpha'}\int \d\tau\bigg[
-\frac{1}{2}\Pi^\mu_+\Bigl\{
\lambda_+\gamma_\mu \theta_-
+\lambda_-\gamma_\mu \theta_+
\Bigr\}
-\Pi^\mu_- \lambda_-\gamma_\mu\theta_-
-\Pi^i_+\Bigl\{
\frac{3}{2}\lambda_-\gamma_i\theta_-
+\frac{1}{2}\lambda_+\gamma_i\theta_+
\Bigr\}
\nn\\&
-\frac{1}{8}\lambda_+\gamma^m\theta_+\,\theta_+\gamma_m\dot\theta_-
+\frac{1}{4}\lambda_+\gamma^i\theta_+\,\theta_+\gamma_i \dot\theta_-
+\frac{1}{8}\lambda_+\gamma^m \dot\theta_+\,\theta_+\gamma_m\theta_-
-\frac{1}{4}\lambda_+\gamma^i \theta_-\,\theta_+\gamma_i\dot\theta_+
\nn\\&
-\frac{11}{24}\lambda_+\gamma^m\theta_-\,\theta_-\gamma_m\dot\theta_-
+\frac{7}{24}\lambda_-\gamma^m\theta_-\,\dot\theta_+\gamma_m\theta_-
+\frac{1}{8}\lambda_-\gamma^m\theta_+\,\theta_+\gamma_m\dot\theta_+
\nn\\&
-\frac{1}{4}\lambda_-\gamma^i\theta_+\,\theta_+\gamma_i\dot\theta_+
+\frac{1}{8}\lambda_-\gamma^m \dot\theta_-\,\theta_+\gamma_m\theta_-
-\frac{1}{4}\lambda_-\gamma^\mu\theta_-\,\theta_+\gamma_\mu\dot\theta_-
~\bigg].
\label{eqn:delta_Q S_b}
\end{align}
It is straightforward to see from \bref{eqn:delta_Q S_0} and \bref{eqn:delta_Q S_b}
that
\begin{align}
\delta_B(S_0+S_b)=\frac{1}{2\pi\alpha'}\int \d\tau\bigg[&
-\Pi^\mu_-\lambda_-\gamma_\mu\theta_-
-\Pi^i_+ \lambda_-\gamma_i\theta_-
\nn\\&
-\frac{1}{3}\lambda_+\gamma^m\theta_-\,\theta_-\gamma_m\dot\theta_-
+\frac{1}{6}\lambda_-\gamma^m\theta_-\, 
\dot\theta_+\gamma_m\theta_-
~\bigg].
\label{eqn:delta S0+Sb}
\end{align}
We have not used any boundary condition so far.
If we impose boundary conditions
$\theta_-=\lambda_-=0$,
\bref{eqn:delta S0+Sb} turns to zero,
as expected.
It is worth noting that 
the $y^i$ dependence has disappeared.
This means that this system is BRST invariant
regardless of the value of $y^i$.
This strongly suggests that
$y^i$ should represent the position of the D$p$-brane.

\section{Coupling to Background Superfields}

In this section, we examine the coupling to background superfields
on a D$p$-brane.
We introduce background superfields
$A_\alpha(\zeta,\bar \eta,\eta),~A_\mu(\zeta,\bar \eta,\eta),~A_i(\zeta,\bar \eta,\eta),~\CW^\alpha(\zeta,\bar \eta,\eta)$
and
\begin{align}
\Fs^\alpha{}_\beta(\zeta,\bar \eta,\eta)\equiv
\delta^\alpha_\beta F_{(0)}
+(\gamma_{mn})^\alpha{}_\beta F^{mn}_{(2)}
+(\gamma_{mnpq})^\alpha{}_\beta F^{mnpq}_{(4)}
\end{align}
where we introduced $\zeta^a=(x^\mu,\theta_+^\alpha)$.
We note $F_{(0)}$, $F_{(2)}^{mn}$ and $F_{(4)}^{mnpq}$
are composed of superfield strength $F_{mn}(\zeta,\bar \eta,\eta)$,
which is consistent to the analysis of
D-brane boundary states \cite{Wyl D brane} by means of
the closed pure spinor superstring.
It is worth noting that our background superfields
are functions of  boundary fermions \cite{SFT boundary fermion},
$\eta_I(\tau)$ and $\bar\eta^I(\tau)$ ($I=1,2,\ldots, q$).
They are needed for the non-abelian generalization of the background superfields.
In \S\ref{sec:DBI},
after quantizing the boundary fermions,
we replace them with Gamma matrices
in $2q$-dimensions.
As a result, background field equations including boundary fermions
turn into the non-abelian version of the background field equations.
Here we note that our background fermions are Dirac fermions.
This plays a key role in proving the nilpotency of the BRST transformation of
boundary fermions
in Appendix \ref{sec:nilpotency}.

In this paper, we introduce the background superfield coupling
\begin{align}
V=
\frac{1}{2\pi\alpha'}\int \d\tau\bigg[&
\dot{\bar \eta}^I\eta_I+
\dot\theta^\alpha_+ A_\alpha(\zeta,\bar \eta,\eta)
+\Pi^\mu_+A_\mu(\zeta,\bar \eta,\eta)
+\Pi^i_- A_i(\zeta,\bar \eta,\eta)
\nn\\&
+\frac{1}{2}\Delta^+_\alpha\CW^\alpha(\zeta,\bar \eta,\eta)
+\frac{1}{4}(N_+)^\beta{}_\alpha(\Fs(\zeta,\bar \eta,\eta))^\alpha{}_\beta
\bigg].
\label{eqn:V}
\end{align}
This background coupling $V$
is manifestly $\epsilon_+$ supersymmetry invariant,
since
$\dot\theta_+$,
$\Pi^\mu_+$,
$\Pi^i_-$,
$\Delta^+$,
$\bar\eta$,
$\eta$
and
\begin{align}
(N_+)^\beta{}_\alpha\equiv& \frac{1}{2} \lambda^\beta_+ w^+_\alpha
\label{eqn:N+}
\end{align}
are $\epsilon_+$ supersymmetry invariant.
Corresponding to \bref{eqn:delta S0+Sb},
we have made the factor $\frac{1}{2\pi\alpha'}$ manifest in $V$.
In order for a background superfield $\Phi\in \{A_\alpha, A_\mu, A_i,\CW^\alpha, {\Fs^{\alpha}}_{\beta} \}$
 to have the correct dimension,
it must be rescaled as $\Phi\to 2\pi\alpha' \Phi$.

We note that
the background superfield
$\Phi$
is hermitian
$\Phi^\dag=\Phi$,
and that $\bar\eta^I=(\eta_I)^\dag$.
These follow from $V^\dag=V$
since 
$(\psi_1\psi_2)^\dag=-\psi_2^\dag \psi_1^\dag$
for Grassmann-odd $\psi_i$.
A superfield $\Phi(\zeta,\bar\eta,\eta)$ may be expanded
in boundary fermions as
\begin{align}
\Phi(\zeta,\bar\eta,\eta)=&
\Phi^{(0)}(\zeta)
-i\bar\eta^I\Phi^J_I(\zeta)\eta_J
+(-i)^2\bar\eta^{I_1}\bar\eta^{I_2}\Phi^{J_1J_2}_{I_1I_2}(\zeta)\eta_{J_1}\eta_{J_2}
+\cdots
\nn\\&
+(-i)^q\bar\eta^{I_1}\cdots\bar\eta^{I_q}\Phi^{J_1\cdots J_q}_{I_1\cdots I_q}(\zeta)\eta_{J_1}\cdots\eta_{J_q}.
\label{eqn:Phi expansion}
\end{align}
The tensor $\Phi^{J_1\cdots J_n}_{I_1\cdots I_n}$
is antisymmetric under exchange of superscripts, $J_1,\ldots,J_n$, and under exchange of subscripts,
$I_1,\ldots,I_n$.
Here we have included $(-i)^n$ in the coefficient of $\Phi^{J_1\cdots J_n}_{I_1\cdots I_n}$.
It follows from 
$\Phi^\dag=\Phi$ that
$(\Phi^{J_1\cdots J_n}_{I_1\cdots I_n})^\dag=\Phi_{J_n\cdots J_1}^{I_n\cdots I_1}$
and  $(\frac{\pa}{\pa\bar\eta^I}\Phi)^\dag=\frac{\pa}{\pa\eta_I}\Phi$.

\section{Supersymmetric DBI Equations 
with Boundary Fermions}
\label{sec:DBI}
 We derive DBI equations for background superfields depending on boundary fermions from the BRST invariance of $S_0+S_b+V$.
The DBI equations obtained in this section
will be promoted to non-abelian DBI equations in the next section.

First of all,
we derive the BRST variation and the time-derivative of the background superfields.
The BRST variation of a superfield $\Phi(\zeta,\bar\eta,\eta)$
 can be written as
\begin{align}
\delta_B \Phi
=\frac{1}{2}\lambda_- \gamma^\mu\theta_- \pa_\mu \Phi
+\lambda^\alpha_+ D_\alpha \Phi
+\delta_B\bar\eta^I  \bar\pa_I \Phi
+\delta_B\eta_I  \pa^I \Phi,
\label{eqn:BRST Phi}
\end{align}
where $D_\alpha=\frac{\pa}{\pa {\theta_+^\alpha}}+\frac{1}{2}(\gamma^\mu\theta_+)_\alpha\pa_\mu$
is the supercovariant derivative on the D$p$-brane worldvolume,
and $\bar\pa_I $ and $\pa^I$ denote $\frac{\pa}{\pa \bar\eta^I}$
and $\frac{\pa}{\pa \eta_I}$, respectively.
Similarly, the time-derivative of $\Phi(\zeta,\bar\eta,\eta)$
is found to be 
\begin{align}
\dot \Phi=\Pi^\mu_+\pa_\mu \Phi +\dot\theta^\alpha_+ D_\alpha \Phi
+\dot{\bar\eta}^I  \bar\pa_I \Phi
+\dot\eta_I  \pa^I \Phi.
\end{align}
After some algebra,
we find the BRST  variation of $V$ as
\begin{align}
\delta_B V=&\frac{1}{2\pi\alpha'}\int \d\tau\bigg[
\Pi^\mu_+\biggl\{
\lambda^\alpha_+( D_\alpha A_\mu-\pa_\mu A_\alpha)
-\lambda_+\gamma_\mu \CW
-\frac{1}{2}\lambda_-\gamma^m\theta_-(\pa_\mu A_m-\pa_m A_\mu) 
\nn\\& 
+\delta_B\bar\eta^I\bar\pa_I A_\mu
+\delta_B\eta_I\pa^I A_\mu
\biggr\}
+\Pi^i_-\biggl\{
\lambda^\alpha_+ D_\alpha A_i
+\frac{1}{2}\lambda_-\gamma^m\theta_- \pa_m A_i
-\lambda_+ \gamma_i \CW
+\delta_B\bar\eta^I\pa_I A_i
\nn\\&
+\delta_B\eta_I\pa^I A_i
\biggr\}
+\Pi^\mu_-\left\{
-\frac{1}{2}\lambda_-\gamma_\mu \CW
-\frac{1}{8}\theta_-\gamma_\mu \Fs \lambda_+
\right\}
+\Pi^i_+\left\{
-\frac{1}{2}\lambda_-\gamma_i \CW 
-\frac{1}{8}\theta_-\gamma_i \Fs\lambda_+
\right\}
\nn\\&
-\frac{1}{2}\Delta^+_\alpha\left\{
\frac{1}{2}\lambda_-\gamma^\mu\theta_- \pa_\mu\CW^\alpha
+\lambda^\beta_+D_\beta \CW^\alpha
-\frac{1}{4}\Fs^\alpha{}_\beta \lambda^\beta_+
+\delta_B\bar\eta^I\bar\pa_I \CW^\alpha
+\delta_B\eta_I\pa^I \CW^\alpha
\right\}
\nn\\&
+\frac{1}{4}(N_+)^\beta{}_\alpha\left\{
\frac{1}{2}\lambda_-\gamma^\mu \theta_- \pa_\mu \Fs^\alpha{}_\beta
+\lambda^\gamma_+ D_\gamma \Fs^\alpha{}_\beta
+\delta_B\bar\eta^I\bar\pa_I \Fs^\alpha{}_\beta
+\delta_B\eta_I\pa^I \Fs^\alpha{}_\beta
\right\}
\nn\\&
+\dot\theta^\alpha_+\left\{-\frac{1}{2}\lambda_-\gamma^m \theta_- (\pa_m A_\alpha-D_\alpha A_m)
-\lambda^\beta_+ (D_\alpha A_\beta+
D_\beta A_\alpha
-\gamma^m_{\alpha\beta}A_m)
-\delta_B\bar\eta^I\bar\pa_I A_\alpha \right.
\nn\\&\left.
-\delta_B\eta_I\pa^I A_\alpha
+\frac{1}{2}(\gamma^m\lambda_-)_\alpha \theta_-\gamma_m\CW
+\frac{1}{4}(\gamma^m\theta_-)_\alpha \lambda_-\gamma_m \CW
+\frac{1}{16}(\gamma^m\theta_-)_\alpha \theta_-\gamma_m\Fs \lambda_+
\right\}
\nn\\&
+\dot\theta^\alpha_-\left\{\frac{1}{2}
(\gamma_m\CW)_\alpha \lambda_+\gamma^m \theta_-
\right\}
+\dot{\bar\eta}^I\left\{
-\delta_B\eta_I
-\lambda^\alpha_+ \pa_I A_\alpha
+\frac{1}{2}\lambda_-\gamma^m\theta_- \bar\pa_I A_m
\right\}
\nn\\&
+\left\{
-\delta_B\bar\eta^I
-\lambda^\alpha_+ \pa^I A_\alpha
+\frac{1}{2}\lambda_-\gamma^m\theta_- \pa^I A_m
\right\}\dot{\eta}_I
\bigg].
\label{eqn:delta V}
\end{align}

\medskip

Gathering
\bref{eqn:delta S0+Sb}
and
\bref{eqn:delta V}
together,
we obtain the BRST variation of $S_0+S_b+V$ as
\begin{align}
\delta_B(S_0+S_b+V)
=\frac{1}{2\pi\alpha'}\int\d\tau\bigg[&
\Pi^\mu_+ X_\mu 
+\Pi^i_- X_i
+\Pi^\mu_- Y_\mu
+\Pi^i_+ Y_i
-\frac{1}{2}\Delta^+_\beta \Lambda^\beta
\nn\\&
+\frac{1}{4}(N_+)^\beta{}_\alpha Z^\alpha{}_\beta
+\dot\theta^\alpha_+\Theta_\alpha^+
+\dot\theta^\alpha_- \Theta_\alpha^-
+\dot{\bar\eta}^I \Xi_I
+\bar \Xi^I \dot\eta_I
~\bigg],
\label{eqn:delta S0+Sb+V}
\end{align}
where $X_m,~Y_m,~\Lambda^\beta,~Z^\alpha{}_\beta,~\Theta_\alpha^\pm,~\Xi_I,~\bar \Xi^I$
are given as follows
\begin{align}
X_m\equiv \, &
\lambda^\alpha_+ (D_\alpha A_m-\pa_m A_\alpha)
-\frac{1}{2}\lambda_-\gamma^n\theta_-(\pa_m A_n-\pa_n A_m)
-\lambda_+\gamma_m \CW
\nn\\&
+\delta_B\bar\eta^I\bar\pa_I A_m
+\delta_B\eta_I\pa^I A_m,
\label{eqn:X}
\\
Y_m\equiv &
-\lambda_-\gamma_m\theta_-
-\frac{1}{2}\lambda_-\gamma_m \CW
-\frac{1}{8}\theta_-\gamma_m \Fs \lambda_+,
\label{eqn:Y}
\\
\Lambda^\alpha\equiv\, &
\frac{1}{2}\lambda_-\gamma^m\theta_- \pa_m\CW^\alpha
+\lambda^\beta_+D_\beta \CW^\alpha
-\frac{1}{4}\Fs^\alpha{}_\beta \lambda^\beta_+
+\delta_B\bar\eta^I\bar\pa_I \CW^\alpha
+\delta_B\eta_I\pa^I \CW^\alpha,
\label{eqn:Lambda}
\\
Z^\alpha{}_\beta\equiv \, &
\frac{1}{2}\lambda_-\gamma^m \theta_- \pa_m \Fs^\alpha{}_\beta
+\lambda^\gamma_+ D_\gamma \Fs^\alpha{}_\beta
+\delta_B\bar\eta^I\bar\pa_I \Fs^\alpha{}_\beta
+\delta_B\eta_I\pa^I \Fs^\alpha{}_\beta,
\label{eqn:Z}
\\
\Theta^+_\alpha\equiv \, &
\frac{1}{6}(\gamma^m\theta_-)_\alpha\, \lambda_-\gamma_m\theta_-
-\frac{1}{2}\lambda_-\gamma^m \theta_- (\pa_m A_\alpha-D_\alpha A_m)
\nn\\&
-\lambda^\beta_+ (D_\alpha A_\beta+D_\beta A_\alpha-\gamma^m_{\alpha\beta}A_m)
+\frac{1}{2}(\gamma^m\lambda_-)_\alpha \theta_-\gamma_m\CW
+\frac{1}{4}(\gamma^m\theta_-)_\alpha \lambda_-\gamma_m \CW
\nn\\&
+\frac{1}{16}(\gamma^m\theta_-)_\alpha\theta_-\gamma_m\Fs \lambda_+
+\delta_B\bar\eta^I\bar\pa_I A_\alpha
+\delta_B\eta_I\pa^I A_\alpha,
\label{eqn:Theta+}
\\
\Theta^-_\alpha\equiv \, &
\frac{1}{3}(\gamma^m\theta_-)_\alpha\,\lambda_+\gamma_m\theta_-
+\frac{1}{2}(\gamma_m\CW)_\alpha \lambda_+\gamma^m \theta_-,
\label{eqn:Theta-}
\\
\Xi_I\equiv&
-\delta_B\eta_I
-\lambda^\alpha_+ \bar\pa_I A_\alpha
+\frac{1}{2}\lambda_-\gamma^m \theta_- \bar\pa_I A_m,
\label{eqn:Xi}
\\
\bar\Xi^I\equiv&
-\delta_B\bar\eta^I
-\lambda^\alpha_+ \pa^I A_\alpha
+\frac{1}{2}\lambda_-\gamma^m \theta_- \pa^I A_m.
\label{eqn:bar Xi}
\end{align}

\medskip

We examine the BRST invariance of $S_0+S_b +V$,
and derive DBI equations of background superfields
with boundary fermions
and boundary conditions on spacetime coordinates.
For this purpose, we will examine the above terms in turn below.
\begin{enumerate}
\item 
$X_i$

This term may be rewritten as
\begin{align}
\delta_B(y_i+A_i)+\lambda_+\gamma_{i}(\theta_{-}+\CW).
\end{align}
To eliminate this term we require
\begin{align}
&\delta_B(y_i+A_i)=0,
\label{eqn:y+A}
\\&
\theta_-^\alpha=- \CW^\alpha.
\label{eqn:theta_-}
\end{align}
We will examine \bref{eqn:y+A} later,
and obtain one of DBI equations with boundary fermions
given in \bref{eqn:BI1}.
The latter equation \bref{eqn:theta_-}
eliminates $\theta_-^\alpha$ from \bref{eqn:delta S0+Sb+V} completely.
Hereafter we understand $\theta_-^\alpha$
as \bref{eqn:theta_-}.
Furthermore,
since \bref{eqn:theta_-} implies
\begin{align}
\dot\theta_-^\alpha=-\dot\CW^\alpha
=-
\left\{
\Pi^\mu_+\pa_\mu \CW^\alpha
+\dot\theta_+^\beta D_\beta \CW^\alpha
+\dot{\bar\eta}^I\bar\pa_I\CW^\alpha
+\dot\eta_I \pa^I \CW^\alpha
\right\}
,
\end{align}
the term  $\dot\theta_-^\alpha \Theta^-_\alpha$
in \bref{eqn:delta S0+Sb+V}  turns to
\begin{align}
-\Bigl\{
\Pi^\mu_+\pa_\mu \CW^\alpha
+\dot\theta_+^\beta D_\beta \CW^\alpha
+\dot{\bar\eta}^I\bar\pa_I\CW^\alpha
+\dot\eta_I \pa^I \CW^\alpha
\Bigr\}\Theta^-_\alpha~,
\label{eqn:dot theta_-}
\end{align}
which contributes to $X_\mu$ in \bref{eqn:X},
$\Theta_\alpha^+$ in \bref{eqn:Theta+},
$\Xi_I$ in \bref{eqn:Xi}
and $\bar\Xi^I$ in \bref{eqn:bar Xi},
respectively.
 
\item
$Y_m$

We find that
$Y_m$ is eliminated by
\begin{align}
\lambda_-^\alpha=-\frac{1}{4}(\Fs)^\alpha{}_\beta\lambda^\beta_+~,
\label{eqn:lambda_-}
\end{align}
which eliminates $\lambda_-^\alpha$ in  \bref{eqn:delta S0+Sb+V} completely.
Hereafter, we understand $\lambda_-^\alpha$ as \bref{eqn:lambda_-}.

\item
$\Xi_I-\bar\pa_I \CW^\alpha \Theta_\alpha^-$
and
$\bar \Xi^I-\pa^I \CW^\alpha \Theta_\alpha^-$

These determine 
the BRST variations of boudary fermions
  as 
\begin{align}
\delta_B \eta_I=&
-\lambda^\alpha_+ \bar\pa_I A_\alpha
+\frac{1}{8}\CW \gamma^m \Fs \lambda_+ \bar\pa_I A_m
+\frac{1}{6}\bar\pa_I \CW \gamma_m\CW\, \lambda_+\gamma^m\CW,
\label{eqn:BRST eta}
\\
\delta_B \bar\eta^I=&
-\lambda^\alpha_+ \pa^I A_\alpha
+\frac{1}{8}\CW \gamma^m \Fs \lambda_+ \pa^I A_m
+\frac{1}{6}\pa^I \CW \gamma_m\CW\, \lambda_+\gamma^m\CW.
\label{eqn:BRST bar eta}
\end{align}
It follows that $(\delta_B\eta_I)^\dag=\delta_B \bar\eta^I$ as expected.
The nilpotency of the BRST transformation of boundary fermions will be shown in
Appendix \ref{sec:nilpotency}.
We find that the BRST variation of a superfield $\Phi$ given in \bref{eqn:BRST Phi} turns to
\begin{align}
\delta_B\Phi=&
\lambda^\alpha_+\left[
\nabla_\alpha \Phi 
+\frac{1}{8}(\CW\gamma^m \Fs)_\alpha\nabla_m\Phi
-\frac{1}{6}(\gamma^m\CW)_\alpha (\gamma_m\CW)_\beta (\CW^\beta,\Phi)
\right]
\equiv \lambda^\alpha_+\hat\nabla _\alpha \Phi.
\label{eqn:hat nabla} 
\end{align}
Here we have introduced covariant derivatives
\begin{align}
\nabla_\alpha=D_\alpha-(A_\alpha,~),~~
\nabla_m=\pa_m+(A_m,~)
\label{eqn:cov derivative}
\end{align}
where we have defined\footnote{
$(-1)^\Omega=-1$ if $\Omega$ is Grassmann-odd,
and  $(-1)^\Omega=1$  if $\Omega$ is Grassmann-even.
}
\begin{align}
(\Phi,\Psi)\equiv& \pa^I \Phi \bar\pa_I\Psi +(-1)^{(\Phi+1)(\Psi+1)}\pa^I\Psi\bar\pa_I\Phi
\nn\\
=&
(-1)^{\Phi+1} \left\{\Phi
\frac{\overleftarrow\pa}{\pa \eta_I}
\frac{\overrightarrow\pa}{\pa \bar\eta^I}
\Psi
-(-1)^{\Phi\Psi}\Psi
\frac{\overleftarrow\pa}{\pa \eta_I}
\frac{\overrightarrow\pa}{\pa \bar\eta^I}
\Phi\right\}.
\label{eqn:( )}
\end{align}
The Jacobi identity which is frequently used below is
\begin{align}
((\Phi,\Psi),\Omega)
+(-1)^{(\Phi+1)(\Psi+\Omega)}((\Psi,\Omega),\Phi)
+(-1)^{(\Omega+1)(\Phi+\Psi)}((\Omega,\Phi),\Psi)
=0.
\label{eqn:Jacobi}
\end{align}
It is easy to show that
\bref{eqn:y+A}
turns out to
\begin{align}
0
=&\lambda^\alpha_+\left[
\nabla_\alpha A_i
+\frac{1}{8}(\CW\gamma^m \Fs)_\alpha\nabla_mA_i
-\frac{1}{6}(\gamma^m\CW)_\alpha (\gamma_m\CW)_\beta (\CW^\beta,A_i)
-(\gamma_i\CW)_\alpha
\right].
\label{eqn:BI1}
\end{align}
This is one of DBI equations with boundary fermions on a D$p$-brane.
As was mentioned below \bref{eqn:delta S0+Sb},
$y^i$ may be regarded as the position of the D$p$-brane.
Accordingly, it is natural to expect
that the equation $\pa_\tau(y_i+A_i)=0$ leads to 
the Dirichlet boundary condition
\begin{align}
&\Pi^i_+-\frac{1}{2}\dot\theta_+\gamma^i \CW
+\Pi^\mu_+ \nabla_\mu A^i
\nn\\&~~
+\Pi^j_- (A_j, A^i)+\dot\theta^\alpha_+ \nabla_\alpha A^i
-\frac{1}{2}\Delta^+_\alpha (\CW^\alpha,A^i)
+\frac{1}{4}N_+^\beta{}_\alpha (\Fs^\alpha{}_\beta,A^i)
=0.
\label{eqn:Dirichlet bc}
\end{align}

\item
$\Lambda^\alpha$

We see that $\Lambda^\alpha=0$ reduces to
\begin{align}
\lambda^\beta_+\left[\hat\nabla_\beta \CW^\alpha -\frac{1}{4}\Fs^\alpha{}_\beta
\right]=0,
\label{eqn:BI2}
\end{align}
where $\hat \nabla$ is defined in \bref{eqn:hat nabla}.
We note that this equation is nothing but
the BRST transformation of \bref{eqn:theta_-}.

\item $Z^\alpha{}_\beta$

We consider the term $(N_+)^\beta{}_\alpha Z^\alpha{}_\beta=\frac{1}{4} \lambda_+^\beta w^+_\alpha Z^\alpha{}_\beta$ in \bref{eqn:Z}.
We note that
\begin{align}
\lambda_+^\beta Z^\alpha{}_\beta=\lambda^\beta_+\delta_B\Fs^\alpha{}_\beta
\end{align}
where   \bref{eqn:hat nabla}
is used.
This term vanishes due to 
the condition which is obtained as
the BRST variation of \bref{eqn:lambda_-}
\begin{align}
\lambda^\beta_+\delta_B\Fs^\alpha{}_\beta=0.
\end{align}
It is worth noting that 
this condition contains the pure spinor constraint.
To see this point,
we further deform the above equation
using  \bref{eqn:hat nabla} and \bref{eqn:BI2}  as
\begin{align}
0=\lambda^\beta_+\lambda^\gamma_+\hat\nabla _\gamma\Fs^\alpha{}_\beta
=4\lambda^\beta_+\lambda^\gamma_+\hat\nabla _\gamma\hat \nabla_\beta \CW^\alpha
=2\lambda^\beta_+\lambda^\gamma_+
\{\hat\nabla _\gamma,\hat \nabla_\beta \}\CW^\alpha.
\label{eqn:nabla nabla}
\end{align}
As shown in \cite{PS open background},
the lowest term which contains neither $\bar \eta$ nor $\eta$
is consistent to 
the pure spinor constraint $\lambda_+\gamma^m\lambda_+
+\lambda_-\gamma^m\lambda_-=0$.
In fact, \bref{eqn:nabla nabla} implies
\begin{align}
0=\lambda^\beta_+\lambda^\gamma_+
\{\hat\nabla _\gamma,\hat \nabla_\beta \}\CW^\alpha\Big|_{\bar\eta,\eta=0}
=(\lambda_+\gamma^m\lambda_++\lambda_-\gamma^m\lambda_-)(\pa_m \CW^\alpha\Big|_{\bar\eta,\eta=0}+\cdots).
\end{align}

\item
$X_\mu-\pa_\mu \CW^\alpha \Theta^-_\alpha$

It is straightforward to see that this is eliminated by
\begin{align}
&
D_\alpha A_\mu
-\pa_\mu A_\alpha
-(A_\alpha,A_\mu)
-\frac{1}{8}(\CW\gamma^m\Fs)_\alpha(\pa_\mu A_m-\pa_m A_\mu
+(A_\mu,A_m))
\nn\\&~~~
+\frac{1}{6}(\gamma^m \CW)_\beta\,(\gamma_m\CW)_\alpha\nabla_\mu\CW^\beta
-(\gamma_\mu\CW)_\alpha
=0.
\label{eqn:BI3}
\end{align}
This is one of DBI equations with boundary fermions on a D$p$-brane.
We note that
\bref{eqn:BI1} and \bref{eqn:BI3} are compactly
expressed as
\begin{align}
&
\hat F_{\alpha m}
-\frac{1}{8}(\CW\gamma^n\Fs)_\alpha \hat F_{mn}
+\frac{1}{6}(\gamma^n \CW)_\beta\,(\gamma_n\CW)_\alpha\nabla_m\CW^\beta
- (\gamma_m\CW)_\alpha
=0.
\label{eqn:BI4}
\end{align}
 Here we have defined $\hat F$ by
 \begin{align}
\hat F_{\alpha m}=&D_\alpha A_m-\pa_m A_\alpha -(A_\alpha, A_m)=-\hat F_{m \alpha},
\label{eqn:hat F alpha m}\\
\hat F_{mn}=&\pa_m A_n -\pa_n A_m+(A_m, A_n).
\label{eqn:hat F mn}
\end{align}

 \item
 $\Theta_\alpha^+
-D_\alpha \CW^\beta \Theta^-_\beta$

To eliminate this term,
we require
\begin{align}
\lambda^\beta_+\bigg[&
\hat F_{\alpha\beta}+\frac{1}{8}(\CW\gamma^m \Fs)_\beta\hat F_{m\alpha}
-\frac{1}{12}(\gamma ^m\CW)_\alpha(\CW\gamma_m \Fs)_\beta
-\frac{1}{6}(\gamma_m\CW)_\gamma (\gamma^m\CW)_\beta \nabla_\alpha \CW^\gamma
\bigg]=0
\label{eqn:7a}
\end{align}
where we have defined
\begin{align}
\hat F_{\alpha\beta}=
D_\alpha A_\beta
+D_\beta A_\alpha
-(A_\alpha,A_\beta)
-\gamma^m_{\alpha\beta} A_m.
\label{eqn:hat F alpha beta}
\end{align}
Eliminating
$\hat F_{m\alpha}$ from \bref{eqn:7a}
by using \bref{eqn:BI4},
we obtain
one of DBI equations with boundary fermions
\begin{align}
\lambda^\beta_+\bigg[&
\hat F_{\alpha\beta} 
-\frac{1}{64}(\CW\gamma^m\Fs)_\beta(\CW\gamma^n\Fs)_\alpha
\hat F_{mn}
\nn\\&
-\frac{1}{6}\Big(
\nabla_\alpha \CW \gamma_m\CW(\gamma^m\CW)_\beta
+\nabla_\beta \CW \gamma_m\CW(\gamma^m\CW)_\alpha
\Big)
\nn\\&
-\frac{1}{36}(\gamma_m\CW)_\gamma(\gamma^m\CW)_\alpha
(\gamma_n\CW)_\delta(\gamma^n\CW)_\beta
(\CW^\delta,\CW^\gamma)
\bigg]=0.
\label{eqn:BI5}
\end{align}

\end{enumerate}

Summarizing, we have shown that
 $\delta_B(S_0+S_b+V)$ in \bref{eqn:delta S0+Sb+V}
 is eliminated completely by
boundary conditions,
\bref{eqn:theta_-}, \bref{eqn:lambda_-} and \bref{eqn:Dirichlet bc},
and by
DBI equations
with boundary fermions,
\bref{eqn:BI2},
\bref{eqn:BI4}
and
\bref{eqn:BI5},
\begin{align}
&
\nabla_\beta \CW^\alpha
+\frac{1}{8}(\CW\gamma^m \Fs)_\beta\nabla_m\CW^\alpha
-\frac{1}{6}(\gamma^m\CW)_\beta (\gamma_m\CW)_\gamma (\CW^\gamma,\CW^\alpha)
 -\frac{1}{4}\Fs^\alpha{}_\beta=0,
 \label{eqn:BI eta 1}
\\&
\hat F_{\alpha m}
-\frac{1}{8}(\CW\gamma^n\Fs)_\alpha \hat F_{mn}
+\frac{1}{6}(\gamma^n \CW)_\beta\,(\gamma_n\CW)_\alpha\nabla_m\CW^\beta
- (\gamma_m\CW)_\alpha
=0,
 \label{eqn:BI eta 2}
\\&
\hat F_{\alpha\beta} 
-\frac{1}{64}(\CW\gamma^m\Fs)_\beta(\CW\gamma^n\Fs)_\alpha
\hat F_{mn}
-\frac{1}{6}\Big(
\nabla_\alpha \CW \gamma_m\CW(\gamma^m\CW)_\beta
+\nabla_\beta \CW \gamma_m\CW(\gamma^m\CW)_\alpha
\Big)
\nn\\&~~~~~
-\frac{1}{36}(\gamma_m\CW)_\gamma(\gamma^m\CW)_\alpha
(\gamma_n\CW)_\delta(\gamma^n\CW)_\beta
(\CW^\delta,\CW^\gamma)
=0
 \label{eqn:BI eta 3}
\end{align}
on a D$p$-brane.

In \S\ref{sec:non-abelian} we will quantize boundary fermions
and obtain non-abelian DBI equations
on coincident D$p$-branes.

\section{Supersymmetric Non-abelian DBI Equations}
\label{sec:non-abelian}
We will quantize boundary fermions, and obtain non-abelian DBI equations
from DBI equations with boundary fermions
given in \bref{eqn:BI eta 1}-\bref{eqn:BI eta 3}.

For this purpose,
 we introduce a pair of canonical momenta conjugate  to the
 boundary fermions  $\bar\eta^I$ and $\eta_I$,
 respectively,
\begin{align}
\pi_I=\frac{\pa L}{\pa \dot{\bar\eta}^I}=\frac{1}{2 \pi\alpha'}\eta_I,~~~
\pi^I=\frac{\pa L}{\pa \dot\eta_I}=0,
\end{align}
where $L=V$ given in \bref{eqn:V}.
For this constrained system,
we define constraints,
$\phi_1\equiv \pi_I-\eta_I/2\pi\alpha'\approx 0$
and $\phi_2\equiv \pi^I \approx 0$. 
The Poisson brackets are given as
 $\{\bar\eta^I,\pi_J\}_\mathrm{PB}=\delta_J^I$
 and
  $\{\eta_I,\pi^J\}_\mathrm{PB}=\delta^J_I$.
Because $\{\phi_1{}_I,\phi_2^J\}_\mathrm{PB}=-\frac{\delta_I^J}{2\pi\alpha'}$
and $\{\phi_1,\phi_1\}_\mathrm{PB}=\{\phi_2,\phi_2\}_\mathrm{PB}=0$,
$\phi_1$ and $\phi_2$  are second-class constraints.
Defining the Dirac bracket by 
$\{F,G\}_\mathrm{DB}\equiv \{F,G\}_\mathrm{PB}-\{F,\phi_\alpha\}_\mathrm{PB}C^{-1\alpha\beta}\{\phi_\beta,G\}_\mathrm{PB}$
where $C_{\alpha\beta}=\{\phi_\alpha,\phi_\beta\}_\mathrm{PB}$,
 we find that\footnote{
For Majorana spinors satisfying $\bar\eta=\eta$,
one obtains $\{\eta^I,\pi_J\}_\mathrm{DB}=\frac{1}{2}\delta_J^I$.
} 
$\{\bar\eta^I,\pi_J\}_\mathrm{DB}=\delta_J^I$,
namely $\{\eta_I,\bar\eta^J\}_\mathrm{DB}=2\pi\alpha' \delta_I^J$,
and that the other brackets become trivial.
It is easy to see that $(\Phi,\Psi)$ defined in \bref{eqn:( )}
can be written as the Dirac bracket
\begin{align}
(\Phi,\Psi)=
\frac{(-1)^{\Phi+1}}{2\pi\alpha'} \{\Phi,\Psi\}_\mathrm{DB}.
\end{align}
Quantization of boundary fermions
is achieved by replacing   the Dirac bracket $\{~,~\}_\mathrm{DB}$
with the anti-commutator multiplied by $\frac{1}{i}$: $\{~,~\}_\mathrm{DB}\to \frac{1}{i}[~,~\}$. 
Through this procedure,
$\{\eta_I,\bar\eta^J\}_\mathrm{DB}=2\pi\alpha' \delta_I^J$
turns into
\begin{align}
\{\eta_I,\bar\eta^J\}=i2\pi\alpha' \delta_I^J.
\label{eqn:CCR}
\end{align}
We note that the factor $i$ is included on the right-hand side.
This is because the anti-commutator of fermions is anti-hermitian
in our notaion\footnote{Of course, if we employ the usual convention in which $(\psi_1\psi_2)^\dag=\psi_2^\dag\psi_1^\dag$,
the anti-commutator \bref{eqn:CCR} turns to  $\{\eta_I,\bar\eta^J\}=2\pi\alpha' \delta_I^J$.
}.

The $\bar\eta$ and $\eta$ satisfying 
the anti-commutator \bref{eqn:CCR}
can be realized with matrices.
To see this, we begin with the Clifford algebra $\{\Gamma^M,\Gamma^N\}=2\delta^{MN}$
in $2q$-dimensions.
Defining $\rho^I$ ($I=1,2,\cdots,q$) from $\Gamma^M$ by
\begin{align}
\rho^I=\frac{1}{2}(\Gamma^I+i \Gamma^{I+q}),~~
\bar\rho^I=\frac{1}{2}(\Gamma^I-i \Gamma^{I+q}),
\end{align}
we see that they satisfy $\{\rho^I,\bar\rho^J\}=\delta^{IJ}$
and $\{\rho^I,\rho^J\}=\{\bar\rho^I,\bar\rho^J\}=0$.
We consider a map from $\eta$ to $\rho$.
First of all, we regard boundary fermions, $\eta$ and $\bar \eta$, as Grassmann-even.
This implies that we multiply each fermion bilinear $\bar\eta^I\eta^J$ by $i$.
The anti-commutator \bref{eqn:CCR} becomes $\{\eta^I,\bar\eta^J\}=2\pi\alpha' \delta^{IJ}$.
Next we rewrite $\eta^I$ and $\bar\eta^J$ as $\sqrt{2\pi\alpha'}\rho^I$ and $\sqrt{2\pi\alpha'}\bar\rho^J$,
respectively.
By this procedure, the background superfield $\Phi(\zeta,\eta,\bar\eta)$ in
\bref{eqn:Phi expansion}
becomes 
\begin{align}
\Phib(\zeta)=&\Phi^{(0)}(\zeta)+2\pi\alpha'\bar\rho^I\Phi_I^J(\zeta)\rho_J
+(2\pi\alpha')^2\frac{1}{(2!)^2}\bar\rho^{IJ}\Phi_{IJ}^{KL}(\zeta)\rho_{KL}
+\cdots\nn\\&
+(2\pi\alpha')^q\frac{1}{(q!)^2}\bar\rho^{I_1\cdots I_q}\Phi_{I_1\cdots I_q}^{J_1\cdots J_q}(\zeta)\rho_{J_1\cdots J_q},
\label{eqn:Phi DBI}
\end{align}
where $\bar\rho^{I_1\cdots I_n}$ is the antisymmetric product of $\bar \rho^{I_1},\ldots,\bar\rho^{I_n}$,
say $\bar\rho^{IJ}=\frac{1}{2}[\bar\rho^I,\bar\rho^J]$,
and so on.
Here and henceforth, the boldface $\Phib$ denotes a matrix.
The superfield strengths in \bref{eqn:hat F alpha beta},
\bref{eqn:hat F alpha m} and \bref{eqn:hat F mn}
turn into
\begin{align}
\Fb_{\alpha\beta}=&
D_\alpha \Ab_\beta
+D_\beta \Ab_\alpha
+\frac{i}{2\pi\alpha'}\{\Ab_\alpha,\Ab_\beta\}
-\gamma^m_{\alpha\beta} \Ab_m,\\
\Fb_{\alpha m}=&D_\alpha \Ab_m-\pa_m \Ab_\alpha +\frac{i}{2\pi\alpha'}[\Ab_\alpha, \Ab_m],
\\
\Fb_{mn}=&\pa_m \Ab_n -\pa_n \Ab_m+\frac{i}{2\pi\alpha'}[\Ab_m, \Ab_n].
\end{align}
The covariant derivatives in \bref{eqn:cov derivative} turn to
\begin{align}
\nabla_\alpha=D_\alpha+\frac{i}{2\pi\alpha'}[\Ab_\alpha,~\},~~
\nabla_m=\pa_m+\frac{i}{2\pi\alpha'}[\Ab_m,~].
\end{align}
As a result,
from the DBI equations in  \bref{eqn:BI eta 1}, \bref{eqn:BI eta 2} and \bref{eqn:BI eta 3},
we obtain the non-abelian DBI equations on coincident D$p$-branes
\begin{align}
&
\nabla_\beta \CWb^\alpha
+\frac{1}{8}(\CWb\gamma^m \Fsb)_\beta\nabla_m\CWb^\alpha
+\frac{i}{12\pi\alpha'}(\gamma^m\CWb)_\beta (\gamma_m\CWb)_\gamma \{\CWb^\gamma,\CWb^\alpha\}
 -\frac{1}{4}\Fsb^\alpha{}_\beta=0,
\label{eqn:DBI 1}
\\&
\Fb_{\alpha m}
-\frac{1}{8}(\CWb\gamma^n\Fsb)_\alpha \Fb_{mn}
+\frac{1}{6}(\gamma^n \CWb)_\beta\,(\gamma_n\CWb)_\alpha\nabla_m\CWb^\beta
- (\gamma_m\CWb)_\alpha
=0,
\label{eqn:DBI 2}
\\&
\Fb_{\alpha\beta} 
-\frac{1}{64}(\CWb\gamma^m\Fsb)_\beta(\CWb\gamma^n\Fsb)_\alpha
\Fb_{mn}
-\frac{1}{6}\Big(
\nabla_\alpha \CWb \gamma_m\CWb(\gamma^m\CWb)_\beta
+\nabla_\beta \CWb \gamma_m\CWb(\gamma^m\CWb)_\alpha
\Big)
\nn\\&~~~~~
+\frac{i}{72\pi\alpha'}(\gamma_m\CWb)_\gamma(\gamma^m\CWb)_\alpha
(\gamma_n\CWb)_\delta(\gamma^n\CWb)_\beta
\{\CWb^\delta,\CWb^\gamma\}
=0.
\label{eqn:DBI 3}
\end{align}

Finally, we make the $\alpha'$ dependence manifest,
so that the background superfields  have correct dimensions.
Rescaling the background superfields as  $\Phi\to 2\pi\alpha'\Phi$, we obtain
\begin{align}
&
\nabla_\beta \CWb^\alpha
 -\frac{1}{4}\Fsb^\alpha{}_\beta
+(2\pi\alpha')^2
\bigg\{\frac{1}{8}(\CWb\gamma^m \Fsb)_\beta\nabla_m\CWb^\alpha
+\frac{i}{6}(\gamma^m\CWb)_\beta (\gamma_m\CWb)_\gamma \{\CWb^\gamma,\CWb^\alpha\}
\bigg\}=0,
\label{eqn:DBI 1'}
\\&
\Fb_{\alpha m}- (\gamma_m\CWb)_\alpha
+(2\pi\alpha')^2
\bigg\{
-\frac{1}{8}(\CWb\gamma^n\Fsb)_\alpha \Fb_{mn}
+\frac{1}{6}(\gamma^n \CWb)_\beta\,(\gamma_n\CWb)_\alpha\nabla_m\CWb^\beta
\bigg\}
=0,
\label{eqn:DBI 2'}
\\&
\Fb_{\alpha\beta} 
-\frac{(2\pi\alpha')^2}{6}\Big(
\nabla_\alpha \CWb \gamma_m\CWb(\gamma^m\CWb)_\beta
+\nabla_\beta \CWb \gamma_m\CWb(\gamma^m\CWb)_\alpha
\Big)
\nn\\&
-(2\pi\alpha')^4\bigg\{
\frac{1}{8^2}(\CWb\gamma^m\Fsb)_\beta(\CWb\gamma^n\Fsb)_\alpha
\Fb_{mn}
-\frac{i}{6^2}(\gamma_m\CWb)_\gamma(\gamma^m\CWb)_\alpha
(\gamma_n\CWb)_\delta(\gamma^n\CWb)_\beta
\{\CWb^\delta,\CWb^\gamma\}
\bigg\}
\nn\\&
=0,
\label{eqn:DBI 3'}
\end{align}
where field strengths and covariant derivatives are given as
\begin{align}
\Fb_{\alpha\beta}=&
D_\alpha \Ab_\beta
+D_\beta \Ab_\alpha
+{i}
\{\Ab_\alpha,\Ab_\beta\}
-\gamma^m_{\alpha\beta} \Ab_m,\\
\Fb_{\alpha m}=&D_\alpha \Ab_m-\pa_m \Ab_\alpha 
+{i}[\Ab_\alpha, \Ab_m],
\\
\Fb_{mn}=&\pa_m \Ab_n -\pa_n \Ab_m
+{i}[\Ab_m, \Ab_n],
\\
\nabla_\alpha=&D_\alpha+{i}[\Ab_\alpha,~\},~~
\nabla_m=\pa_m+{i}[\Ab_m,~].
\end{align}
Correspondingly the expansion \bref{eqn:Phi DBI}
changes to
\begin{align}
\Phib(\zeta)=&\Phi^{(0)}(\zeta)+
\bar\rho^I\Phi_I^J(\zeta)\rho_J
+2\pi\alpha'\frac{1}{(2!)^2}\bar\rho^{IJ}\Phi_{IJ}^{KL}(\zeta)\rho_{KL}
+\cdots\nn\\&
+(2\pi\alpha')^{q-1}\frac{1}{(q!)^2}\bar\rho^{I_1\cdots I_q}\Phi_{I_1\cdots I_q}^{J_1\cdots J_q}(\zeta)\rho_{J_1\cdots J_q}.
\label{eqn:Phi DBI scaled}
\end{align}
Here $\Phi^{(0)}$ is scaled 
as $\Phi^{(0)}\to 2\pi\alpha' \Phi^{(0)}$
so as to reproduce the abelian case \cite{HS DBI,PS open background}.

Our result is consistent with the non-abelian DBI equations on coincident D9-branes
derived by using the superembedding approach  in \cite{boundary fermion} .
In the abelian case,
our result  reduces to one derived in \cite{HS DBI}\footnote{
Our result also reduces to one derived by using the superembedding approach for a single D9 brane \cite{Ker D9 DBI}
(see also \cite{Cayley image}).}.
If we expand our result to the second order in boundary fermions,
the result given in \cite{PS open background} is reproduced. 

\smallskip

We comment on the limit $\alpha'\to0$.
It is straightforward to see that the non-abelian DBI equations \bref{eqn:DBI 1'}-\bref{eqn:DBI 3'} reduce 
to\footnote{We note that \bref{eqn:SYM 1} can be derived also from the Bianchi identity $2\nabla_{(\alpha}\Fb_{\beta)m}-\gamma^n_{\alpha\beta}\Fb_{nm}=0$
and \bref{eqn:SYM 2}.
}
\begin{align}
\nabla_\beta\CWb^\alpha
+\frac{1}{4}
(\gamma^{mn})^\alpha{}_\beta\Fb_{mn}
&=0,
\label{eqn:SYM 1}\\
\Fb_{\alpha m}-(\gamma_m \CWb)_\alpha&=0,
\label{eqn:SYM 2}\\
\Fb_{\alpha\beta}&=0.
\label{eqn:SYM 3}
\end{align}
Here we used the fact that $\Fsb^\alpha{}_\beta=-(\gamma^{mn})^\alpha{}_\beta\Fb_{mn}$
in the limit $\alpha'\to 0$.
These are nothing but the SYM equations.
Furthermore, it is obvious from  \bref{eqn:Phi DBI scaled}
that a background superfield
reduces to
\begin{align}
\Phib(\zeta)=&\Phi^{(0)}(\zeta)+\bar\rho^I\Phi_I^J(\zeta)\rho_J,
\end{align}
in the limit $\alpha'\to0$.
This implies that the gauge symmetry of the SYM is U(1)$\times$SU($q$)
when $\tr \Phi_I^J=0$ is imposed.
As a result, we have shown that non-abelian DBI equations reduce to the SYM equations in the limit 
$\alpha'\to 0$,
as expected.

\section{Summary and Discussion}
We derived supersymmetric non-abelian D-brane equations from
the open pure spinor superstring.
First, a boundary action $S_b$ is introduced so that 
half of 32 supersymmetries of the pure spinor superstring action $S_0$
may be preserved
without imposing any boundary condition for a D$p$-brane.
In this paper, we introduced background superfields on a D$p$-brane
which are  functions of boundary fermions.
Including the coupling $V$ to these background superfields,
we examined the condition for the BRST invariance of $S_0+S_b+V$.
We solved this condition to obtain the supersymmetric DBI equations for the background superfields
as well as boundary condition on the spacetime coordinates.
By quantizing boundary fermions, these supersymmetric DBI equations with boundary fermions
are promoted to supersymmetric {non-abelian} DBI equations,
\bref{eqn:DBI 1'}, \bref{eqn:DBI 2'} and \bref{eqn:DBI 3'}.
We have shown that in the limit $\alpha'\to 0$, these equations surely reduce to SYM equations as expected.
Furthermore, the BRST variations of boundary fermions,  which are determined by the BRST invariance of $S_0+S_b+V$,
are  shown to be nilpotent.

By expanding
\bref{eqn:DBI 1'}, \bref{eqn:DBI 2'} and \bref{eqn:DBI 3'} to the second order in boundary fermions,
the result given in \cite{PS open background} which contains contributions to non-abelian equations at the lowest order in $\alpha'$ is reproduced. 

The supersymmetric non-abelian DBI equations on coincident D9-branes
are derived in \cite{boundary fermion}
using the superembedding approach.
They extended the superspace to include boundary fermions,
and so find some difficulty in quantizing boundary fermions only.
In this paper, boundary fermions are treated separately,
and so the  canonical quantization is achieved straightforwardly.

The superembedding approach was developed to multiple D0 brane system in \cite{Bandos D0}.
It is interesting to pursue these issues from the open string point of view.

We note that the boundary fermions we introduced are Dirac fermions.
Accordingly, the supersymmetric DBI equations with boundary fermions enjoy  a global U$(q)$ symmetry.
This symmetry is utilized to show the nilpotency of the BRST variation of the boundary fermions  in Appendix A.
Extending our model to one with Majorana boundary fermions
is left as an interesting problem.

It is interesting to derive equations of motion of background fields
from our supersymmetric non-abelian DBI equations together with the Bianchi identities.
These equations of motion should be derived from the D$p$-brane action
which is composed of the DBI action and the Wess-Zumino action.
In the limit $\alpha'\to 0$, 
our DBI equations are shown to reduce to the SYM equations.
At the next $\alpha'^2$ order, we expect to obtain equations which correspond to the DBI action.
There are some attempts to derive equations at the  $\alpha'^4$ order,
but there seems to be a difficulty to obtain equations in a supersymmetric form,
except for the $p=3$ case in which the harmonic superspace is utilized \cite{DHHK0305}.

\section*{Acknowledgments}
The authors would like to thank Takanori Fujiwara,
Yoshifumi Hyakutake and Hiroshi Kunitomo for useful comments.
The authors thank the Yukawa Institute for Theoretical Physics at Kyoto University. Discussions during the YITP workshop YITP-W-22-09 on "Strings and Fields 2022" were useful to complete this work.
This work was supported by JSPS KAKENHI Grant Number JP21K03566, 
by JST, the establishment of university
fellowships towards the creation of science technology innovation, Grant Number JPMJFS2105, 
and by the Sasakawa Scientific Research Grant from The Japan Science Society.

 \appendix
 
 \section*{Appendix}
 
\section{On BRST-variations of Boundary Fermions}
\label{sec:nilpotency}

We comment on the BRST transformation of $\bar \eta$ and $\eta$ given in  \bref{eqn:BRST eta} and \bref{eqn:BRST bar eta}.
These are determined
by the BRST invariance of $S_0+S_b+V$.
Since background superfields may be expanded as in 
\bref{eqn:Phi expansion},
we find that
these can be expressed as
\begin{align}
\delta_B\eta_I =O_I{}^J\eta_J,~~~
\delta_B \bar\eta^I =\bar\eta ^JO_J{}^I,
\label{eqn:delta eta}
\end{align}
where $O$ is Grassmann-odd. 
Since $(\delta_B \eta_I)^\dag=\delta_B \bar\eta^I$
follows from  \bref{eqn:BRST eta} and \bref{eqn:BRST bar eta},
$O$ is found to be anti-hermitian, $O^\dag=-O$.
Acting $\delta_B$ on \bref{eqn:delta eta}, we obtain
\begin{align}
\delta_B^2\eta_I
=(\delta_{B}O_I{}^J-O_I{}^KO_K{}^J)\eta_J,~~~
\delta_B^2\bar\eta^I
=-\bar\eta^J(\delta_{B}O_J{}^I-O_J{}^KO_K{}^I).
\label{eqn:nil}
\end{align}
For  the BRST transformation to be nilpotent,
these must vanish.
We show that these terms can be eliminated by gauge transformations.

Note that the total action $S_0+S_b+V$ is invariant under the global U($q$) transformation\footnote{Note that
$\Phi(\zeta,\bar\eta,\eta)$  is invariant under U($q$).
Coefficients of $\bar\eta$ and $\eta$
in \bref{eqn:Phi expansion}
transform as
$\Phi^{(0)}\to \Phi^{(0)}$,
$\Phi_I^J  \to U_I{}^K\Phi_K^L U^\dag_L{}^J $,
and so on.
}
\begin{align}
\eta_I\to\eta_I'=U_I{}^J\eta_J,~~
\bar\eta^I\to\bar\eta'^I =\bar\eta^J U^\dag_J{}^I,
\end{align}
where $I,J=1,2,\ldots,q$.
In order to gauge this symmetry, we introduce a gauge field $a(\tau)$ and replace the kinetic term $\dot{\bar\eta}^I \eta_I$
with $-\bar\eta^I D\eta_I\equiv-\bar\eta^I(\delta_I^J\pa_\tau+a_I{}^J)\eta_J$.
The infinitesimal gauge transformation law is
\begin{align}
\delta \eta_I=i\theta(\tau)_I{}^J \eta_J,~~\delta \bar\eta^I=-i\bar\eta^J \theta(\tau)_J{}^I,~~
\delta a_I{}^J=-i\dot\theta(\tau)_I{}^J+i[\theta(\tau),a]_I{}^J
\end{align}
where the gauge parameter $\theta$ is hermitian, $\theta^\dag=\theta$.
We require that the gauge field $a$ is invariant under the BRST transformation.
This implies that the analysis on the BRST invariance of $S_0+S_b+V$
is not affected by the inclusion of $a$,
except that $\pa_\tau$ is replaced with $D$.
By including the gauge transformation,
\bref{eqn:nil} becomes
\begin{align}
\delta_B^2\eta_I
=&\left(\delta_{B}O_I{}^J-O_I{}^KO_K{}^J+i\theta(\tau)_I{}^J\right)\eta_J,\\
\delta_B^2\bar\eta^I
=&-\bar\eta^J\left(\delta_{B}O_J{}^I-O_J{}^KO_K{}^I+i\theta(\tau)_J{}^I\right).
\end{align}
As $(i\theta)^\dag=-i\theta$,
we can gauge fix $\theta(\tau)$ so that $\delta_B^2\eta_I=\delta_B^2\bar\eta^I=0$
if  $(\delta_{B}O-OO)^\dag=-(\delta_{B}O-OO)$.
We show that this is the case.
First, it is obvious that $(O_I{}^KO_K{}^J)^\dag=-O_J{}^KO_K{}^I$ as $O$ is Grassmann-odd.
Next we examine $\delta_B O$,
which may be expressed from \bref{eqn:hat nabla} as
\begin{align}
\delta_BO=&
\lambda^\alpha_+
\nabla_\alpha O
+\frac{1}{8}\CW\gamma^m \Fs\lambda_+\nabla_mO
-\frac{1}{6}\lambda_+\gamma^m\CW (\gamma_m\CW)_\beta (\CW^\beta,O).
\end{align}
Nothing the fact that $O$ is anti-hermitian,
we see that $\nabla_\alpha O$ and $\nabla_mO$ and $(\CW^\beta,O)$
are anti-hermitian. 
This implies that $(\delta_B O)^\dag=-\delta_B O$.
As a result, we have shown that $(\delta_{B}O-OO)^\dag=-(\delta_{B}O-OO)$,
so that we can make $\delta_B^2\eta_I=\delta_B^2\bar\eta^I=0$ by using the U($q$) gauge transformation.

\end{document}